\documentclass{article}
\usepackage[utf8]{inputenc}
\usepackage{amssymb}
\usepackage{graphicx}
\usepackage{amsmath}
\usepackage{bm}
\usepackage{subfigure}
\usepackage{siunitx}

\usepackage{soul}
\usepackage{color}
\newcommand{\vinodh}[1]{{\color{black}#1}}

\newcommand{\etal}{\textit{et al.}}

\title{Simulating the non-linear interaction of relativistic electrons and tokamak plasma instabilities: Implementation and validation of a fluid model}
\author{V.~Bandaru, M.~Hoelzl, F.J.~Artola, G.~Papp, G.T.A.~Huijsmans}
\date{December 2018}

\begin{document}

\maketitle

\begin{abstract}
For the simulation of disruptions in tokamak fusion plasmas, a fluid model describing the evolution of relativistic runaway electrons and their interaction with the background plasma is presented. The overall aim of the model is to self-consistently describe the non-linear coupled evolution of runaway electrons (REs) and plasma instabilities during disruptions. In this model, the runaway electrons are considered as a separate fluid species in which the initial seed is generated through the Dreicer source, that eventually grows by the avalanche mechanism (further relevant source mechanisms can easily be added). Advection of the runaway electrons is considered primarily along field lines, but also taking into account the $\bm{E}\times\bm{B}$ drift. The model is implemented in the non-linear magnetohydrodynamic (MHD) code JOREK based on Bezier finite-elements, with current-coupling to the thermal plasma. Benchmarking of the code with the one-dimensional runaway electron code GO is done using an artificial thermal quench on a circular plasma. As a first demonstration, the code is applied to the problem of an axisymmetric cold vertical displacement event in an ITER plasma, revealing significantly different dynamics between cases computed with and without runaway electrons. Though it is not yet feasible to achieve fully realistic runaway electron velocities close to the speed of light in complete simulations of slowly evolving plasma instabilities, the code is demostrated to be suitable to study various kinds of MHD-RE interaction in MHD-active and disruption relevant plasmas.

%\matthias{An outlook to further planned improvements of %the model and the numerical treatment is given.}

\end{abstract}

\section{Introduction}
\label{section1}

In tokamak plasmas, a disruption refers to the sudden loss of plasma confinement due to large scale magnetohydrodynamic instabilities. During the `thermal quench' (TQ) phase of the disruption, the plasma loses its thermal energy within a short timescale (\vinodh{$\sim 0.5$ms to $0.7$ms for most tokamaks, with the time increasing with the minor radius}) due to the stochastization of the magnetic field, cooling down the plasma by several orders of magnitude to temperatures of the order of 10eV. This increases the electrical resistivity ($\eta$) of the plasma significantly, leading to the decay of the plasma current on the resistive timescale, referred to as the current quench (CQ). The decay of the current gives rise to a large toroidal electric field that can accelerate \vinodh{suprathermal} electrons to relativistic velocities and energies of the order of a few tens of MeV. Such electrons, known as runaway electrons (REs) would eventually carry all the toroidal plasma current by the end of the current quench, which is estimated to be a large fraction ($\sim 60\%$) of the predisruption current in fusion relevant devices~\cite{Hender:2007}. Uncontrolled loss of REs can lead to deep melting of plasma facing components and unacceptably long machine downtimes. This is the general motivation for the study of the formation, interaction with the background plasma, and losses of runaway electrons.

In view of their very low collisionality, in principle, a kinetic representation would be apt to model runaway electron behaviour accurately. However, due to the prohibitive computational overhead, REs are often modelled via passive particle tracing, such as in the simulations of Izzo \etal~\cite{Izzo:2011} and Sommariva \etal~\cite{Sommariva:2018, Sommariva2:2018}. In these simulations, the electromagnetic field history obtained from a disruption simulation without REs is used to track the motion of a few thousand runaway electrons seeded randomly in the plasma volume. Although such simulations yield useful insights into the transport, generation, and deconfinement of REs in the stochastic field, the back reaction of the REs on the background plasma is unaccounted for. A fluid model for REs complements the particle tracer model, by consistently treating the coupling of the REs with MHD. Studies of REs interacting with the resistive kink modes have been conducted using an RE fluid model by Cai \etal~\cite{Cai:2015} using the M3D code and Matsuyama \etal~\cite{Matsuyama:2017} in the context of the spectral code EXTREM, which is limited to cylindrical plasmas. 

In this paper, we present a runaway electron fluid model that is implemented in the JOREK code~\cite{Huijsmans:2007, Czarny:2008}. JOREK is a fully-implicit 3D non-linear MHD code based on 2D Bezier finite elements in the poloidal plane and a Fourier decomposition in the toroidal direction. The code can handle realistic X-point tokamak geometries and is routinely used for MHD simulations of edge localized modes (ELMs) and disruptions. The free-boundary extension of JOREK, referred to as JOREK-STARWALL~\cite{Merkel:2015, Hoelzl:2012}, also includes the electromagnetic response of the structures outside the plasma, such as the vacuum vessel, central solenoid, field coils etc.

The newly implemented runaway electron fluid is coupled to the MHD primarily through the RE currents in the evolution of poloidal flux. Interaction of REs with MHD also occurs through the ion momentum equation, that defines the ideal MHD equilibrium state of the plasma in the presence of REs. Numerical stabilization using the Taylor-Galerkin approach (TG2) enables achieving higher parallel advection velocities for the runaway electrons for a given timestep.

The paper is organised as follows:~The RE fluid model and its coupling to MHD is described in section \ref{section2}, followed by benchmarking and numerical tests in section \ref{section3}. In section \ref{section4}, we describe the application of the model to simulate a cold vertical displacement event of an ITER plasma, that is followed by summary and conclusions in section \ref{section5}.

\section{Runaway electron fluid model and coupling with reduced MHD}
\label{section2}
In our model, the runaway electrons are considered as a separate fluid species that interacts with the single-fluid representation of the background plasma consisting of the thermal ions and electrons. In addition, it is assumed that all the REs move at the speed of light along the field-parallel direction with the $\bm{E}\times\bm{B}$ drift superimposed. The velocity of REs is denoted by $\bm{v_r}$ and is given by
\begin{equation}
\bm{v_r} = c\frac{\bm{B}}{B} + \frac{\bm{E}\times \bm{B}}{B^2},
\end{equation}
where $c$ is the speed of light, $\bm{E}$ is the electric field, $\bm{B}$ denotes the magnetic field and $B=|\bm{B}|$ is the magnitude of the magnetic field. \vinodh{The curvature drift of the REs is neglected here for the sake of simplicity. Unlike the thermal plasma, the curvature drift of REs can be important in the context of equilibrium and MHD behaviour. Nevertheless, the extent to which the neglect of curvature drifts can affect the solutions is not yet fully clear and will be considered in the future.} The considered JOREK physics model uses a reduced MHD formulation wherein the magnetic and the electric field are expressed through the poloidal flux $\psi$ and the electric potential $u$ respectively as
\begin{equation}
\begin{split}
\bm{B} &= R^{-1} \left[ \nabla \psi \times \bm{e_{\phi}} + F_0 \bm{e_{\phi}} \right], \\
\bm{E} &= -F_0 \nabla u - R^{-1} \partial_t \psi.
\end{split}
\end{equation}
Here, $\bm{e_{\phi}}$ is the unit vector in the toroidal direction, $R$ is the major radial coordinate and $F_0$ is a constant. In this framework, the $\bm{E}\times\bm{B}$ drift velocity is expressed as 
\begin{equation}
\frac{\bm{E}\times \bm{B}}{B^2} \approx \frac{-F_0 \nabla u \times R^{-1}F_0 \bm{e_{\phi}}}{\left(F_0/R\right)^{2}} = -R \left(\nabla u \times \bm{e_{\phi}} \right).
\end{equation}
For the parallel advection of the RE density, due to the numerical difficulty in advecting at the speed of light, a downscaling factor ($f$) is used when necessary such that the parallel advection velocity $c_a$ is given by
\begin{equation}
c_a = fc,  \quad 10^{-2} \le f \le 1.
\end{equation}
This is presently needed to deal with the large separation between the true parallel advection timescale $\sim \SI{e-8}{\second}$ and the timescale of MHD changes $\tau_\mathrm{MHD}\sim \SI{e-4}{\second}$, that is relevant for example to tearing modes. This downscaling is justified for a number of problems of interest that does not involve stochastic fields, since an advection velocity of $\sim \SI{e6}{\meter \second^{-1}}$ is already significantly larger than $\tau_\mathrm{MHD}$  and ensures that the RE density redistributes nearly uniformly over the flux surfaces on the MHD timescale. For example, it was shown in Ref.~\cite{Matsuyama:2017} that the non-linear evolution of the $\left(1,1\right)$ kink mode becomes insensitive to the RE advection velocity as long as it is significantly larger than the Alfv{\'e}n speed. 
Using the above considerations, the advection of the RE number density $n_r$ can be expressed as
\begin{equation}
\begin{split}
&-\nabla \cdot \left(\vinodh{\bm{v_{r,a}}} n_r\right) \\ &=  -\nabla \cdot \left[ \frac{c_a n_r}{B} \bm{B} + n_r \frac{\bm{E}\times\bm{B}}{B^2} \right]
\\  & = -c_a \bm{B} \cdot \nabla \left(\frac{n_r}{B}\right) +  \nabla \cdot \left[n_r R \left(\nabla u \times \bm{e_{\phi}} \right)\right] 
\\ & = -\frac{c_a}{B R} \left[ \left[n_r,\psi\right] + \frac{F_0}{R} \frac{\partial n_r}{\partial \phi}  \right] + 
\frac{c_a n_r}{B^2 R} \left[ \left[B,\psi\right] + \frac{F_0}{R} \frac{\partial B}{\partial \phi}  \right] \\ & + R\left[n_r,u\right] + 2n_r \partial_z u,
\end{split}
\end{equation}
\vinodh{where $\bm{v_{r,a}}$ is the velocity used for RE advection,} $z$ denotes the vertical coordinate and the Poisson bracket operator is defined such that $\left[n_r,u\right] = \partial_R \left(n_r\right) \partial_z u - \partial_z n_r \partial_R u$.

However, such a downscaling of the parallel advection velocity is not  fully realistic when the magnetic field is stochastic, especially when one is interested in the radially outward transport of REs, which would be underestimated. For such circumstances, we have the option to mimic the fast parallel advection of REs in a stochastic field through a parallel diffusion term $\nabla\cdot\left(D_{\parallel,r}\nabla_\parallel n_{r}\right)$ instead of the parallel advection term, where $D_{\parallel,r}$ is the parallel diffusivity of REs. In a stochastic field, the radial location of the field lines evolves in a diffusive way when tracing them. Particles moving along the field lines will therefore also experience a radial diffusion with time, which reduces radial gradients of the particle density. The perpendicular motion by drifts will effectively also have a diffusive character by moving particles from one field line to another one (or a different location along the same). Whether a group of particles is moving along the field lines convectively or diffusively does not make a significant difference since the physical radial diffusion of the particles can be modelled by both ways when an appropriate parallel diffusion coefficient is chosen, which can be determined either by field line tracing or by analytical estimates. The net effect of annihilating gradients of $n_r$ along field lines on a fast timescale remains the same. This ensures that the parallel diffusion model can effectively reproduce the features of RE transport that affects MHD in a stochastic field. For instance, particles can be lost from a stochastic field line region while they stay confined in island remnants, leading to an effective helical current perturbation affecting MHD stability. The effective parallel RE diffusivity in a stochastic tokamak plasma can be estimated to be $D_{\parallel,r}  \sim cL_c/\pi^2$, where the length scale is chosen to be the auto-correlation length $L_c=\pi R$ \cite{Rechester:1978}. This leads to an estimate for the parallel diffusivity $D_{\parallel, r}\sim Rc/\pi \sim 10^{8}$-$\SI{e9}{\meter^2 \second^{-1}}$. Assuming that the perpendicular diffusivity $D_{\perp, r}$ would be about the same magnitude as that due to turbulent diffusion ($D_{\perp, r} \sim \SI{1}{\meter^2 \second^{-1}}$), we obtain a $D_{\parallel, r}/D_{\perp, r} \sim 10^8$-$10^9$. Such values of the ratio of parallel to perpendicular diffusivities for REs can be  treated well in JOREK, as it is done with the parallel thermal diffusivity. A further improvement of the numerical scheme is presently being considered, which should allow to resolve even higher anisotropy.

The total current density in the plasma $\bm{j}$ is decomposed into the thermal and runaway electron components as
\begin{equation}
\bm{j} =  \bm{j_\mathrm{th}} + \bm{j_r}, \quad \bm{j_r} = -en_r \bm{v_r},
\end{equation}
where $\bm{j_\mathrm{th}}$ is the thermal electron current density, $\bm{j_r}$ is the RE current density and $e$ represents the electron charge. Primary generation (or seeding) of REs due to diffusion in the velocity space (Dreicer mechanism) is modelled as a volumetric source term $S_p$ given by Connor\etal~\cite{Connor:1975}
\begin{equation}
\begin{split}
 S_p &= \left(0.21+0.11Z_e\right) n \nu_\mathrm{ee} \epsilon_d ^{-\frac{3}{16}\left(1+Z_e\right)}
 e^{\left( -\frac{1}{4}\epsilon_d^{-1} - \left(1+Z_e\right)^{1/2} \epsilon_d^{-1/2} \right) } \\ & \times e^{\left[ -\frac{T_e}{m_e c^2} \left(\frac{1}{8}\epsilon_d^{-2} + \frac{2}{3}\left(1+Z_e\right)^{1/2} \epsilon_d^{-3/2} \right) \right] },
 \end{split}
\end{equation}
where $Z_e$ is the effective ion charge, $\nu_\mathrm{ee}$ the electron-electron collision frequency, $m_e$ and $T_e$ are the electron mass and temperature respectively, and $\epsilon_d=E_{\parallel}/E_{d}$ is the ratio of the parallel electric field to the Dreicer electric field. Here, the Dreicer electric field~\cite{Dreicer:1959} is given by 
\begin{equation}
E_d = \frac{n_e e^3 \ln{\Lambda}}{4 \pi \epsilon_0^2 T_e},
\end{equation}
where $\ln \Lambda$ is the Coulomb logarithm and the parallel electric field is defined as $E_\parallel = \left(\bm{E} \cdot \bm{B}\right)/B$. The amplification of the seed REs through large angle knock-on collisions is modelled using the Rosenbluth-Putvinski model~\cite{Rosenbluth:1997} as
\begin{equation}
\begin{split}
S_s &= n_r \nu_\mathrm{fp} \frac{\epsilon_c -1}{\ln{\Lambda}} \sqrt{\frac{\pi \varphi}{3\left(Z_e+5\right)}} \\ & \times \left( 1 - \frac{1}{\epsilon_c} + \frac{4 \pi \left(Z_e + 1 \right)^{2}}{3 \varphi \left(Z_e + 5\right) \left(\epsilon_c^2 + 4/\varphi^2 -1\right)}  \right) ^{-1/2},
 \end{split}
\end{equation}
where $S_s$ is the volumetric secondary source of REs, $\nu_\mathrm{fp}$ is the Fokker-Planck collision frequency, $\varphi = \left( 1 + 1.46 \sqrt{\epsilon} + 1.72 \epsilon\right)^{-1}$ is the neoclassical function with $\epsilon= r/R$ being the aspect ratio, and $\epsilon_c = E_{\parallel}/E_c$ with the critical electric field $E_c$ given by
\begin{equation}
E_c =  \frac{n_e e^3 \ln{\Lambda}}{4 \pi \epsilon_0^2 m_e c^2}.
\end{equation}
As the presently available fluid approximations to hot-tail generation are limited in applicability \vinodh{(\cite{Smith:2008},~\cite{Martinsolis:2017})}, we do not consider the hot-tail generation mechanism in our model at present, while a term describing it can be added later on or an ad-hoc seed distribution can be initialized. Also, it must be noted that, while the secondary RE generation occurs at timescales close to the resistive timescale $\tau_\mathrm{res}=\mu_0 L^2/\eta \sim \SI{0.1}{\second}$ for a $\SI{10}{\electronvolt}$ plasma, the Dreicer generation is a relatively faster process occuring at a timescale $\tau_\mathrm{Drecier} \sim 10^{-5}$-$\SI{e-6}{\second}$ in a typical tokamak plasma. Hence the Dreicer generation has a much stronger dynamical coupling to the MHD than the secondary RE generation. We now turn to the coupling of REs to the momentum equation.

It can be easily seen that the presence of REs leads to an additional term in the single-fluid momentum equation, equivalent to $e n_{r} \bm{E} - \bm{j_r}\times\bm{B}$. This arises due to the (albeit small) involvement of the RE population in maintaining charge neutrality.
However, the $\bm{j}_{r}\times\bm{B}$ term can be simplified as follows.
\begin{align}
\bm{j_r}\times\bm{B} &= -e n_{r} \bm{v_r}\times\bm{B} \\
\notag &= -e n_{r} \left[ \frac{c\bm{B}}{B} + \frac{\bm{E}\times\bm{B}}{B^2}  \right] \times \bm{B} \\
\notag &= -e n_{r} \left[ \frac{\left( \bm{E}\cdot\bm{B}\right)\bm{B} } {B^2} - \bm{E}\right] \\
\notag &= e n_{r} \bm{E} - \frac{e n_{r}}{B^2}\left(\bm{E}\cdot\bm{B}\right)\bm{B} \\
\notag &= e n_{r} \bm{E} - \frac{e n_{r}}{B}E_{||}\bm{B}.
\end{align}
Therefore,
\begin{align}
e n_{r} \bm{E} - \bm{j_r}\times\bm{B} &= \frac{e n_{r}}{B}E_{||}\bm{B}.
\end{align}
Using the above, the single-fluid momentum equation for the background plasma becomes
\begin{align}
\label{momentum_equation}
\rho \frac{d\bm{v}}{dt} &= \frac{e n_{r}}{B}E_{||}\bm{B} + \bm{j} \times \bm{B} - \nabla p - \bm{v}S_{\rho},
\end{align}
where $\rho$ is the ion mass density, $p$ is the total pressure from the ion and thermal electron components, $S_{\rho}$ is the mass density source and $\bm{v}$ is the plasma ion fluid velocity. It is important to note that the correction term due to REs can be of the same order of magnitude as the material derivative of the velocity (L.H.S) after the thermal quench due to a large parallel electrical resistivity. The ion fluid velocity is decomposed as
\begin{align}
\bm{v} &= -R\nabla u \times \bm{e_\phi} + v_\parallel \frac{\bm{B}}{B} - \frac{\nabla p \times \bm{B}}{neB^2}.
\end{align}
In order to obtain evolution equations for $u$ and $v_\parallel$, the momentum equation \eqref{momentum_equation} is projected respectively by the operators $\nabla \phi \cdot \nabla \times \left[R^2 \left(..\right)\right]$ and $\bm{B}\cdot\left(..\right)$, where $\nabla \phi = \bm{e_\phi}/R$.
Using the above considerations, the full set of equations coupling reduced MHD with the RE fluid model in JOREK can be written in normalized units as below (see the appendix for details of normalization). For simplicity, the same variable names have been retained for the normalized variables.

\begin{align}
\label{psi_equation}
\frac{1}{R^2} \partial_t \psi &= \frac{\eta\left(T\right)}{R^2} \left(j -cn_{r}\frac{F_0}{BR}\right)  - \frac{\eta_h}{R}\nabla^2\left(\frac{j-cn_{r}F_0/\left(BR\right)}{R}\right) \notag \\& - \frac{F_0}{R^2} \partial_\phi u  - \frac{1}{R} \left[u,\psi\right] + \frac{\tau_{IC}}{\rho} \frac{F_0^2}{R^2 B^2}  \left(\frac{F_0}{R^2} \partial_\phi p + \frac{1}{R} \left[p,\psi\right]\right)
\end{align}
\begin{align}
\label{u_equation}
&\nabla\cdot\left[\rho R^2\nabla_\perp \frac{\partial u}{\partial t}\right] \notag \\ & = \frac{1}{2R}\left[R^2 |\nabla_\perp u|^2,R^2 p\right] + \frac{1}{R}\left[R^4 \rho\omega,u\right] + \frac{1}{R} \left[\psi,j\right]  \notag \\ & - \frac{F_0}{R^2}\partial_\phi j -\frac{1}{R}\left[R^2,\rho T\right] + R\mu_\perp\left(T\right)\nabla^2\omega + R\mu_{\perp,h}\left(T\right)\nabla^4\omega \notag \\&
-\frac{1}{RB}\left[ n_{r} E_{||} \left(\partial_{RR}\psi + \partial_{ZZ}\psi \right) +   
\partial_R \psi \partial_R \left(n_r E_\parallel \right) + \partial_Z \psi \partial_Z \left(n_r E_\parallel \right) \right] \notag \\&
+ \tau_\mathrm{IC}\left( R^3 \left[w_e,p\right] + R^2\nabla\cdot\left(\partial_z p \nabla_\perp u \right) \right) \notag \\& - \tau_\mathrm{IC} R^3 \left[\partial_{RZ}u\left(\partial_{RR}p-\partial_{ZZ}p\right)-\partial_{RZ}p\left(\partial_{RR}u-\partial_{ZZ}u\right)
\right] \notag \\&- \nabla\cdot\left[R^2 \nabla_\perp u \left( S_\mathrm{part} + \left(1-n_c\right)\left( \rho \rho_n S_\mathrm{ion}\left(T\right) - \rho^2\alpha_\mathrm{rec}\left(T\right)\right)\right)\right]
\end{align}
\begin{align}
\label{jequation}
j &= \Delta^{*}\psi
\end{align}
\begin{align}
\omega &= \nabla\cdot\nabla_{\perp}u
\end{align}
\begin{align}
& \partial_t \rho \notag \\ &= R\left[\rho,u\right] + 2\rho\partial_z u + \frac{\rho}{R}\left[\psi,v_\parallel\right] + \frac{v_\parallel}{R}\left[\psi,\rho\right] \notag \\& - \frac{F_0}{R^2}\left(v_\parallel\partial_\phi \rho + \rho\partial_\phi v_\parallel\right)  + 2\tau_\mathrm{IC}\partial_z p + \nabla\cdot\left(D_\parallel\nabla_\parallel\rho + D_\perp\nabla_\perp\rho\right) \notag \\&  + D_{\perp,h}\nabla^{4}\rho + S_\mathrm{part} + \rho \rho_n S_\mathrm{ion}\left(T\right) - \rho^2\alpha_\mathrm{rec}\left(T\right)
\end{align}
\begin{align}
\label{p_equation}
&\partial_t \left(\rho T\right) \notag \\ &= R\left[\rho T,u\right] - {v_\parallel} \left(\frac{1}{R}\left[\rho T,\psi\right) + \frac{F_0}{R^2}\partial_\phi\left(\rho T\right)\right] +   2\gamma\rho T\partial_z u \notag \\ & -\gamma\rho T \left(\frac{1}{R}\left[v_\parallel,\psi\right]+\frac{F_0}{R^2}\partial_\phi v_\parallel\right) +  \nabla\cdot\left(\kappa_\perp\nabla_\perp T + \kappa_\parallel\nabla_\parallel T\right) \notag \\ & +\kappa_{h}\nabla^4 T  + S_q   + \frac{\left(\gamma -1\right) }{R^2}\eta\left(T\right){\left(j-cn_{r}\frac{F_0}{BR}\right)}^2  - \rho L_\mathrm{bg} \notag \\& - \xi_\mathrm{ion}\rho\rho_n S_\mathrm{ion}\left(T\right)  - \rho \rho_n L_\mathrm{lines}\left(T\right) - \rho^2 L_\mathrm{brem}\left(T\right) 
\end{align}
\begin{align}
\label{vpar_equation}
&\rho B^2 \partial_t v_\parallel \notag \\ &= \frac{B n_{r}}{R}E_{||} -\frac{\rho F_0}{2R^2} \partial_\phi \left({v_\parallel}^2 B^2\right) - \frac{\rho}{2R}\left[{v_\parallel}^2 B^2,\psi\right]  + \frac{1}{R} \left[\psi,\rho T\right] \notag \\ & - \frac{F_0}{R^2} \partial_\phi \left(\rho T\right) + \mu_\parallel \left(T\right) B^2 \nabla^2v_\parallel  + \mu_{\parallel,h} \left(T\right) B^2 \nabla^4v_\parallel \notag   \notag \\& -{v_\parallel} B^2 \left( S_\mathrm{part} +  \left(1-n_c\right)\left[\rho\rho_nS_\mathrm{ion}\left(T\right)-\rho^2\alpha_\mathrm{rec}\left(T\right)\right]\right),
\end{align}
\begin{align}
\partial_t \rho_n &= \nabla\cdot\left(D_n\nabla\rho_n\right) - \rho\rho_n S_\mathrm{ion}\left(T\right) + \rho^2\alpha_\mathrm{rec}\left(T\right) + \vinodh{S_\mathrm{MMI} }
\end{align}

\begin{align}
\label{nr_equation}
&\partial_t n_r  \notag \\ &=  \left(i_D -1\right) \left[-\frac{c_a}{B R} \left( \left[n_r,\psi\right] + \frac{F_0}{R} \partial_\phi n_r  \right)  + 
\frac{c_a n_r}{B^2 R} \left( \left[B,\psi\right] + \frac{F_0}{R} \partial_\phi B  \right)\right] \notag \\ & + R\left[n_r,u\right] + 2n_r \partial_z u + \nabla\cdot\left(i_D D_{\parallel,r}\nabla_\parallel n_{r} + D_{\perp, r}\nabla_\perp n_{r}\right) + D_{r,\perp,h} \nabla^4n_{r} \notag \\&
+ S_p + S_s,
\end{align} 
where the parallel electric field is treated parametrically and is given by
\begin{align}
\label{epar_equation}
E_{||}\left(T,j,n_{r}\right) &=   -\frac{\eta(T)F_0}{BR^2}\left(j - cn_{r}\frac{F_0}{BR}\right) \notag \\& - \frac{\tau_\mathrm{IC}}{\rho} \frac{F_0^3}{B^3 R^2} \left( \frac{1}{R} \left[p,\psi\right] + \frac{F_0}{R^2} \partial_\phi p \right). 
\end{align} 
Equations \eqref{psi_equation} to \eqref{epar_equation} form a closed set for the unknown scalar variables $\left[\psi \quad u \quad j \quad \omega \quad \rho \quad T \quad v_\parallel \quad \rho_n \quad n_r \quad E_\parallel\right]$, where $j$ and $\omega$ represent the toroidal component of \vinodh{the} total current density and vorticity respectively, $v_\parallel$ is the field parallel velocity component and $\rho_n$ is the mass density of neutrals. In addition, $\tau_\mathrm{IC}$ is the diamagnetic factor, $\gamma=5/3$ and the operator in equation \eqref{jequation} is defined as $\nabla^{*}\left(\cdot\right)=\left[R\partial_R\left(R^{-1}\partial_R \left(\cdot\right) \right) + \partial_{ZZ} \left(\cdot\right)\right]$. The variable sets $\left[\mu_\perp \quad \mu_\parallel\right]$, $\left[D_\perp \quad D_\parallel\right]$, $\left[\kappa_\perp \quad \kappa_\parallel\right]$ denote the perpendicular and parallel components of the viscosity, mass diffusion coefficient and thermal diffusivity respectively, whereas the variables with the subscript `$h$' denote hyperdiffusion coefficients in the respective equations. Furthermore, the terms $S_\mathrm{part}$, $S_\mathrm{ion}$, $\alpha_\mathrm{rec}$ denote ion mass density sources due to fuelling, ionisation and recombination respectively, whereas $S_\mathrm{MMI}$ denotes the neutral mass density source due to massive material injection. Finally, the terms $S_q$, $L_\mathrm{lines}$, $L_\mathrm{bg}$, $L_\mathrm{brem}$ represent respectively the thermal energy sources/sinks from plasma heating, line radiation, background radiation and bremsstrahlung. In the applications shown in this paper, the above sources are however not used. The prefactor $i_D$ in equation~\eqref{nr_equation} indicates a Boolean integer, where $i_D=1$ denotes the use of parallel RE diffusion instead of parallel advection. Note that the RE number density $n_r$ enters the MHD model through equations \eqref{psi_equation}, \eqref{u_equation}, \eqref{p_equation} and \eqref{vpar_equation}. Further details of the reduced MHD model in JOREK with neutrals and \vinodh{massive material injection} can be found in Fil \etal~\cite{Fil:2015} and Nardon \etal~\cite{Nardon:2017}. JOREK also has models for impurity massive material injection, which is not shown here. All the governing equations are implemented in the weak form in JOREK. Details of the normalization of the variables appearing in equations \eqref{psi_equation} to \eqref{epar_equation} that are important in the context of this paper, along with the expressions for the normalized sources $S_p$ and $S_s$ in equation \eqref{nr_equation} are given in the appendix. Note that the magnetic field $\bm{B}$, spatial dimensions and the poloidal flux $\psi$ remain unnormalized. 

\section{Numerical stabilization and benchmarking}
\label{section3}
\subsection{Taylor-Galerkin stabilization for RE advection}
As mentioned earlier, parallel advection of REs occurs at the speed of light which is at least three to four orders of magnitude larger than the timescale of MHD changes that we are interested in. In spite of the downscaling of the parallel advection velocity used in our code in some cases, numerical instabilities  leading to the contamination of the solution may arise already with a $c_a \sim \SI{e6}{\meter\second^{-1}}$ when using time steps suitable for the MHD dynamics. This issue is often encountered when Galerkin schemes are used to treat strong advection phenomena, which is also the case in JOREK. To stabilize the scheme for RE advection, we use the approach of Taylor-Galerkin (TG2) stabilization~\cite{Donea:1984,Donea:1987}. Such a stabilization is also used in JOREK for the $v_{\parallel}$ and $\bm{E}\times\bm{B}$ ion advection terms, in which case the velocities are of the order of $\SI{e4}{\meter \second^{-1}}$. In treating the RE advection terms with TG2, in effect we add the following term to the right hand side of the discretized version of equation~\eqref{nr_equation}
\begin{equation}
 - \frac{f_\mathrm{TG}}{2}\frac{\Delta t}{2} \nabla \cdot \left[ \left(R \left[n_r, u\right]\bm{v_D} - \frac{c}{B R} \left[ \left[n_r,\psi\right]  + \frac{F_0}{R} \frac{\partial n_r}{\partial \phi}  \right]  c \frac{\bm{B}}{B} \right) \right],
\end{equation}
before conversion to the weak form.  This gives rise to an effective numerical diffusion, which stabilizes the advection operator. Here, $f_\mathrm{TG}$ is a weighting constant of order $0.1$ associated with the stabilization term.

\begin{figure}[!h]{}
\centering
\includegraphics[width=0.6\textwidth]{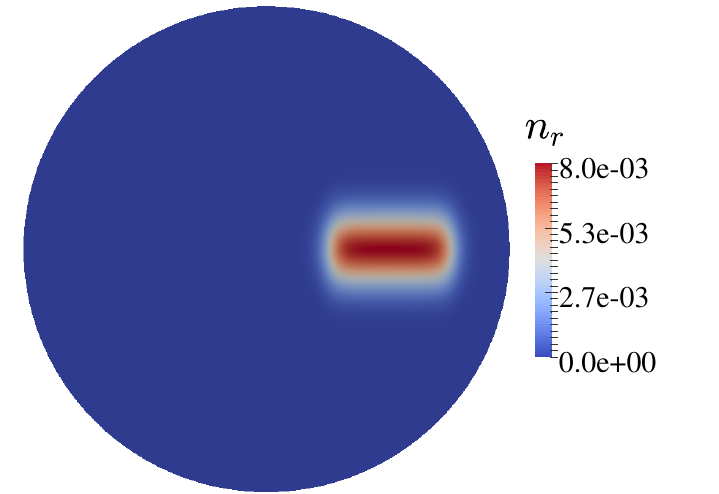}]
\caption{Runaway electron number density $n_r$ at time $t=0$.}
\label{figure:nre_at_t0_pure_advection}
\end{figure}

\begin{figure}[!h]
\subfigure[]{
\centering
\includegraphics[width=0.5\textwidth]{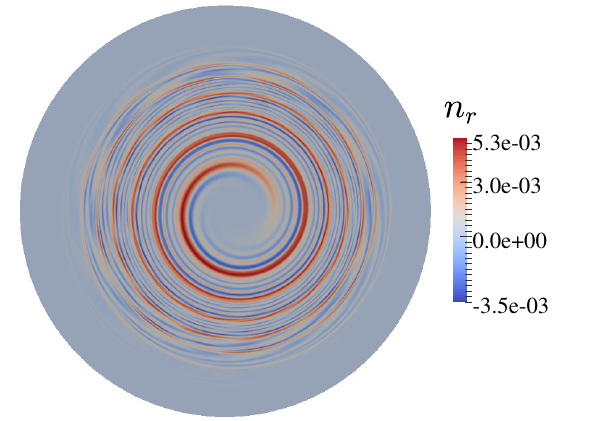}
\label{figure:nr_without_tgnum}
}
\subfigure[]{
\centering
\includegraphics[width=0.5\textwidth]{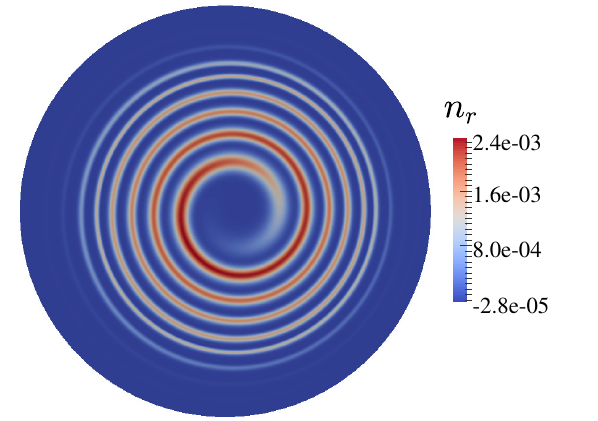}
\label{figure:nr_with_tgnum}
}
\caption{Runaway electron number density $n_r$ for the pure advection test case after time $t=2$ (normalized) with a parallel advection velocity $c_a=c$ ($f=1$) with a) No stabilization and b) TG2 stabilization, $f_\mathrm{TG}=0.25$.} 
\label{figure:imp_with_tgnum}
\end{figure}

In order to demonstrate the effectiveness of the TG2 stabilization in enabling relatively high RE advection velocities, we describe here a test case of pure parallel advection of an initial spatial distribution of REs in a circular plasma in static equilibrium. In other words, RE generation sources and $\bm{E}\times\bm{B}$ advection are not included and we keep the background electromagnetic field fixed in time. The initial distribution of RE density is given by $n_r\left(t=0\right) = f_1\left(Z\right) f_2\left(R\right)$, where $f_1$ and $f_2$ are given by
\begin{align}
f_1 \left(Z\right) &= \frac{b}{w\sqrt{2 \pi}} e^{-\frac{Z^2}{2w^2}}, \notag \\
f_2 \left(R\right) &= \tanh \left(x\right) + \tanh \left( 20-x \right) - 1, \quad R \ge R_\mathrm{axis},
\end{align} 
with $x=20 \left(R-R_\mathrm{axis} \right)$ and $b$ being a constant. The corresponding contour plot of the initial RE density distribution is shown in Fig.~\ref{figure:nre_at_t0_pure_advection}. Such a poloidally localized $n_r$ distribution
is not typically encountered in experiments, and would constitute a much more stringent test case for parallel RE advection than a case in which $n_r$ is approximately uniform within a closed flux surface. The values of the safety factor $q$ for the equilibrium varies between $q=1.3$ at the axis to $q=3.6$ at the plasma edge. Such a radial variation of the rotational transform leads to a gradual stretching of the $n_r$ distribution with time due to parallel advection, evolving into a spiral-shaped distribution in the poloidal plane, in a way that conserves the number of REs between any two closed flux surfaces. This process does not lead to a steady state and gives rise to increasingly (radially) localized distribution with time. This was simulated using a $70\times80$ poloidal grid resolution with fixed boundary condition for $n_r$ and with various values of $c_a$. The $n_r$ distribution obtained for $c_a=c$ and $\Delta t = 2\times 10^{-3}$ after a normalized time $t=2$ is shown in Fig.~\ref{figure:imp_with_tgnum} (Note that the solution of this problem for a fixed value of $c_a t$ is independent of the specific values of $c_a$ and $t$ used). Use of TG2 (see Fig.~\ref{figure:nr_with_tgnum}) clearly leads to a significant improvement in the solution as compared to a completely unphysical solution obtained without TG2 as is shown in Fig.~\ref{figure:nr_without_tgnum}. That the application of TG2 still leads to a conservative solution was confirmed from several time traces of $\int_{\psi_{N,1}}^{\psi_{N,2}} n_r dV$, where $\psi_{N,2}-\psi_{N,1}=0.1$. Here, $\psi_N$ refers to the normalized poloidal flux defined as $\psi_N = \left(\psi-\psi_\mathrm{axis}\right)\left(\psi_\mathrm{bnd}-\psi_\mathrm{axis}\right)^{-1}$, where $\psi_\mathrm{axis}$ and $\psi_\mathrm{bnd}$ are the flux values at the plasma axis and the domain boundary respectively. For example, within the band $0.1 \le \psi_N \le 0.2$, the maximum deviation of the integral without and with TG2 applied was $0.02\%$ and $0.06\%$ respectively. \vinodh{Both the error values in the integral are rather small and the one without TG2 is in fact smaller. However, it is important to note that unlike the case with TG2, the simulations without TG2 leads to large negative values for the number density $n_r$, of the same order of magnitude as the maximum positive value of $n_r$ in the solution. In addition, the solution for $n_r$ without TG2 displays spurious and sharp localized spikes in the spatial distribution, contrary to the relatively smooth solutions obtained with TG2.}

 Although very advantageous, it can however be observed that the use of TG2 cannot obviate the use of small timesteps for $c_a \sim c$. Hence, for practical applications, it would be feasible to achieve RE advection velocities up to $c_a \sim 10^{-2}c$. A plausible way to practically access advection velocities close to the speed of light would be the use of a multirate method, in which, for each timestep used to evolve the MHD system without equation~\eqref{nr_equation}, several much smaller timesteps are used to evolve $n_r$ through equation~\eqref{nr_equation}. Multirate numerical schemes are typically designed for and used in systems where the physical processes of interest have a timescale seperation atmost $\sim 10$. In the present context, such a method would lead to a loss of the fully-implicit coupling between the MHD system and the RE fluid density. In addition, given that the timescale seperation in the MHD-RE problem considered here is $\sim 10^4$, it is very likely that a multirate scheme would not be feasible. Further investigations regarding this numerical challenge are left for future work, since the present model already allows to investigate many physically relevant processes as discussed in the following \vinodh{sections}.
 
 \subsection{Thermal electron to RE current conversion}

\begin{figure}[!h]
\subfigure[]{
\includegraphics[width=0.5\textwidth]{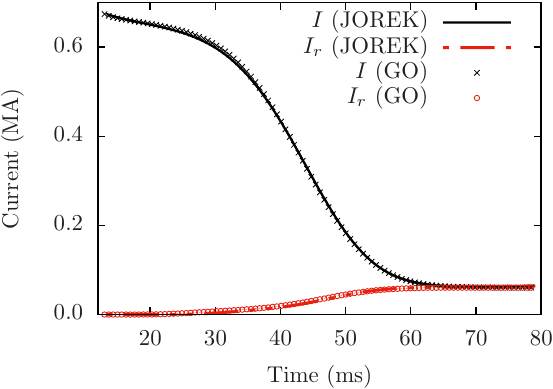}
\label{fig:jorek_vs_go}
}
\subfigure[]{
\includegraphics[width=0.5\textwidth]{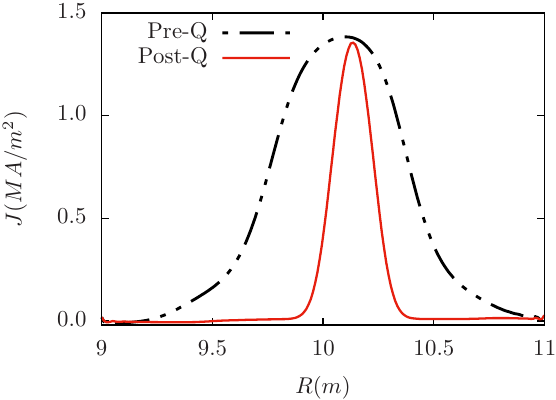}
\label{figure:re_peaking}
}
\caption{a) Time evolution of the total plasma current $I$ and the RE current $I_{r}$ during the current quench phase. b) Midplane current density profiles before and after the current quench obtained from JOREK, showing a relatively peaked RE current profile.} 
\end{figure}

The RE fluid model in JOREK was benchmarked with the one-dimensional runaway electron code GO~\cite{Papp:2013} for the conversion of thermal electron current to RE current. This was done by triggering an artificial thermal quench in a large aspect ratio circular plasma with major radius $R=10$m and minor radius $a=1$m. At the initial state, the plasma is in equilibrium with a plasma current $I_p=\SI{0.67}{\mega\ampere}$, on-axis toroidal magnetic field $B_{\phi,0}=\SI{1}{\tesla}$ and with respective central temperature and density of $T_0=\SI{1.7}{\kilo \electronvolt}$ and $n_0=\SI{1e20}{\meter^{-3}}$. The initial parallel electric field is purely Ohmic and the resistivity of the plasma is assumed to vary as $\eta\left(T\right) = \eta_0 \left(T/T_0\right)^{-1.5}$ with the initial central resistivity $\eta_0=\SI{1.1e-7}{\ohm\meter}$. The plasma is then quenched (thermally) by imposing a large perpendicular thermal diffusivity $\kappa_{\perp} =100 \kappa_{\perp,\mathrm{eq}}$, as compared to the thermal diffusivity at equilibrium $\kappa_{\perp,\mathrm{eq}}$. This leads to a drop in the core temperature of the plasma to about $\SI{25}{\electronvolt}$ in a time of about $\SI{60}{\milli\second}$, which is much slower than typical tokamak thermal quench times $\sim 1$ms. This case is run in JOREK using a $51\times24$ poloidal grid size in an axisymmetric setting without any toroidally asymmetric modes, with no neutrals in the plasma and with fixed boundary conditions for all the variables. An RE advection velocity $c_a=10^{-3}c$ was used. Thresholds were set in this case to initiate the Drecier generation when $E_\parallel/E_D \geq 0.01$ and the avalanching when $E_\parallel/E_c \geq 1.7$. Especially the avalanching threshold helps in avoiding the unphysical behaviour that can potentially occur through the amplification of numerical noise near the plasma edge, where the ratio $E_\parallel/E_c$ is the highest. As the focus is on the verification of the implementation of RE dynamics, the temperature profile evolution is not self-consistently calculated in GO, but rather taken as input from JOREK. The current quench in this case occurs at the same timescale as that of the thermal quench. Figure~\ref{fig:jorek_vs_go} shows a very good match of the result obtained \vinodh{with} GO and JOREK for the evolution of the total and runaway electron currents. 
%Such numerical issues near the plasma edge have also been encountered in some MHD simulations with runaway electrons by the users of the M3D code
The simulations also show the centrally peaked runaway electron current profile, which is often observed in experiments. This can be seen in the pre-quench and post-quench current density profiles shown in Fig.~\ref{figure:re_peaking}. The peaked profile occurs due to the resistive diffusion of the parallel electric field towards the centre, where the RE formation is most effective~\cite{Eriksson:2004}. 

\subsection{Linear growth of the internal kink mode with REs}
\begin{figure}[!h]
\subfigure[]{
\includegraphics[width=0.5\textwidth]{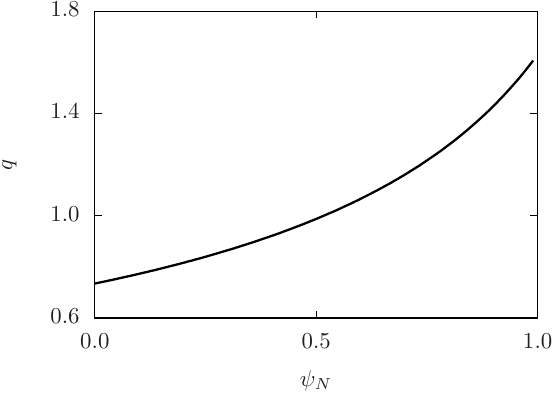}
\label{fig:qprofile_kink}
}
\subfigure[]{
\includegraphics[width=0.5\textwidth]{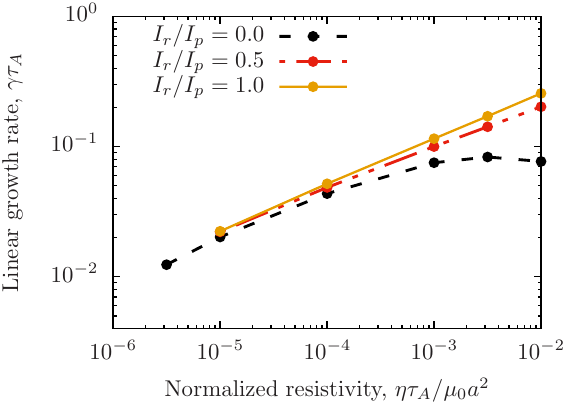}
\label{fig:LGR_vs_eta_kink}
}
\caption{a) Profile of the safety factor $q$ considered for the internal kink case. b) Linear growth rate of the resistive internal kink $\left(1,1\right)$ mode as a function of the normalized resistivity at various initialized fractions of runaway current. Here, Alfv{\'e}n time $\tau_A=a\sqrt{\mu_0 \rho_0}/B$ and $\gamma$ is the growth rate in SI units.}
\end{figure}

This study is aimed at the verification of the qualitative behaviour of the linear growth of the internal kink mode, when a part of the plasma current is assumed to be carried by runaway electrons instead of thermal electrons. This is done by considering again a large aspect ratio circular plasma ($R=\SI{10}{\meter}$ and $a= \SI{1}{\meter}$) in a fixed-boundary static equilibrium ($\bm{v}=0$) with parameters $B_{\phi,0}=\SI{1}{\tesla}$, $I_p=\SI{0.31}{\mega\ampere}$ and a low on-axis temperature $T_0=\SI{48}{\electronvolt}$. The equilibrium is $\left(m=1,n=1\right)$ kink unstable with the $q=1$ surface within the plasma as shown in Fig.~\ref{fig:qprofile_kink}. We now initialize the runaway electron density so as to have qualitatively the same profile as the total current and carry a fraction of the total plasma current. Three different cases with RE current fraction $I_{r}/I_p=0,0.5,1.0$ have been considered.
At time $t=0$, a small perturbation is applied to follow its linear growth. Computations were performed using a $80\times46$ poloidal grid with local radial-refinement to resolve the kink-mode. Thermal and mass diffusivities along with all the sources (including RE generation) were set to zero and the resistivity was assumed to be temperature independent and spatially constant. Runaway electron advection is also neglected in this case. Figure~\ref{fig:LGR_vs_eta_kink} shows the linear growth rate of the internal kink mode as a function of normalized resistivity (inverse Lundquist number $S^{-1}$) for the various fractions of RE current considered. It can be observed that an increase in the RE current fraction leads to a gradual recovery of the low resistivity scaling $S^{-1/3}$ even at large values of the normalized resistivities. This occurs primarily due to the reduced effective resistivity in the presence of runaway electrons, which have practically negligible collision frequency. In other words, when the RE current fraction is increased, the region outside the resistive layer tends towards the ideal MHD limit in which the low-resistivity analytical scaling is valid. Our results show a qualitatively similar behaviour to those observed by Matsuyama \etal~\cite{Matsuyama:2017} in a similar but not identical case.

\section{ITER vertical displacement event (VDE) simulation with runaway electrons}
\label{section4}

\begin{figure}[!h]{}
\centering
\includegraphics[width=0.7\textwidth]{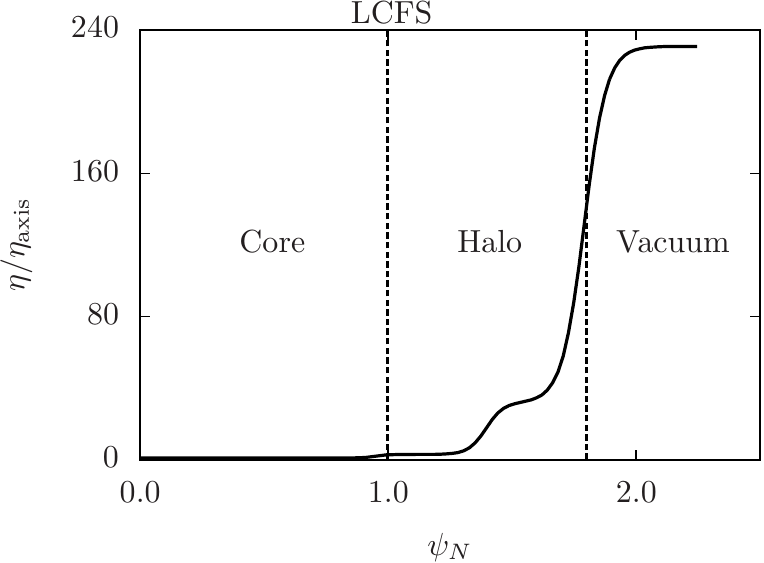}
\caption{Normalized resistivity $\eta/\eta_\mathrm{axis}$ as a function of the normalized poloidal magnetic flux $\psi_N$. }
\label{figure:eta_profile}
\end{figure}

We now apply the model to simulate an axisymmetric cold VDE with REs in an ITER plasma. ITER VDE simulations will be investigated in detail in a separate publication by Artola \etal. VDE simulations with JOREK based on an NSTX equlibrium have been benchmarked recently with M3D-C\textsuperscript{1} \cite{Krebs:2019}. One specific simulation is used in the present paper to demonstrate the capabilities of our RE model and to give an example for a relevant physics study the model can be applied to. In \vinodh{particular}, we consider the case of a non-stochastic, post thermal-quench ITER plasma, that is subjected to an axisymmetric vertical motion with the simultaneous generation of runaway electrons. At time $t=0$, the plasma is in a state of static (velocity $\bm{v}=0$) free-boundary equilibrium, with a small seed density of runaway electrons with a Gaussian spatial distribution given by
\begin{equation}
n_r \left(t=0\right) = \frac{d}{w\sqrt{2\pi}} e^{-\frac{\psi_N^2}{2w^2}},
\end{equation}
where $\psi_N$ is the normalized poloidal flux, $w$ is the distribution width and $d$ is a constant. The current carried by the runaway electron seed $I_{r}\left(t=0\right)\sim10^{-3} I_p$. Furthermore, at the initial state, the total plasma current $I_p=\SI{14.5}{\mega\ampere}$ and the toroidal magnetic field at the plasma axis $B_{\phi,0}=\SI{4.8}{\tesla}$. The density of the plasma is assumed to be spatially uniform and time independent, with $n_e=\SI{5e19}{\meter^{-3}}$. The velocity field is assumed to consist of only the $\bm{E}\times\bm{B}$ drift. The resistivity $\eta$ is considered to be a function of poloidal flux rather than temperature. The resistivity profile used is shown in Fig.~\ref{figure:eta_profile}, where $\eta_\mathrm{axis}=\SI{1.24e-4}{\ohm \meter}$, \vinodh{which corresponds to the Spitzer resistivity at $T=\SI{2.35}{\electronvolt}$. The halo region in the figure refers to the region in the scrape-off layer (SOL) wherein the resistivity is sufficiently low that significant currents can exist. With the given resistivity profile we do not intend to  accurately model the evolution of halo currents in ITER, but rather is used as a simple test case for the runaway model. Note that the considered plasma resistivity gives a current quench (CQ) time of 10 ms, which is not representative of the expected CQ time in ITER mitigated disruptions ($50$ms $< \tau_\mathrm{CQ, ITER} < 150$ms). The chosen profile is therefore rather arbitrary, but has numerical advantages such as a broad halo region, (up to $50\%$ of the normalized poloidal flux) and a small jump of a factor $3$ in the resistivity from the inner LCFS to the SOL. The evolution of the halo region temperature and its associated resistivity during VDEs in disruptions is still not well established. The prediction of the halo temperature and the halo width requires to accurately simulate different effects such as plasma radiation, impurity sources and transport, parallel transport and ohmic heating which are out of the scope of this paper.  Measurements in Alcator C-mod~\cite{tinguely:2017} indicate that the halo region width varies between $15\%-60\%$ of the normalized poloidal flux, which is consistent with our resistivity profile.

Due to the large vertical motion, we need to take into account the far SOL region and therefore solve the resistive MHD equations as well. After the defined halo region, a very large resistivity is imposed (``vacuum region") in order to remove the stabilizing effect of the eddy currents that would be induced there if the resistivity were too low.}

An RE advection velocity $c_a=10^{-4}c$ and a small value of the diffusion \vinodh{coefficient} for RE density, $D_{\perp,r}=10^{-8}$ (normalized units) was used for numerical reasons. A constant viscosity $\mu=\SI{3.9 e-4}{\kg \meter^{-1} \second^{-1}}$ was used. The effect of the \vinodh{poloidal} field coils, central solenoid, and the vacuum vessel on the plasma response is \vinodh{modelled} by the use of JOREK-STARWALL. This means that the external structures are not explicitly included in the computational domain, but rather treated in a numerically efficient way by the use of non-local (integral) boundary conditions for $\psi$ through the Green's function formalism. Hence the problem domain is limited to the region until the first plasma-wall interface. The configuration for external conductors used in the present simulations is similar to that in Artola \etal~\cite{Artola:2018} \vinodh{and Gribov \etal~\cite{Gribov:2015}, but without the vertical stabilization (VS) coils. The active coils modeled include six coils comprising the central solenoid (CS) and six poloidal field (PF) coils. The passive conducting structures modeled are the outer triangular support (OTS), the divertor inboard rail (DIR) and the stainless steel vacuum vessel. The vacuum vessel is two-layered with each layer having a thickness of $\SI{6}{\centi\meter}$}. The boundary conditions for $u$, $\omega$ and $n_{r}$ are however fixed in time. Although not fully realistic, these boundary conditions are expected to provide useful estimates of the effect of RE growth on the MHD dynamics during most part of the VDE. More realistic boundary conditions for the velocity field and the runaway current are presently being developed. 

Simulations were run with a radial-poloidal grid resolution of $170\times240$ points with the evolution equations for $\rho$, $T$, $v_{\parallel}$ and $\rho_n$ switched off. In \vinodh{particular}, two different simulations were performed, each of them with and without the generation of runaway electrons: a purely axisymmetric simulation ($n=0$) for a total time of $\SI{10.6}{\milli\second}$ and a simulation with two non-axisymmetric toroidal Fourier modes ($n=0,1,2$) for a total time of $\SI{8.6}{\milli\second}$. The simulations without runaway electrons are referred to as `baseline' in the remainder of the text. Due to the relatively high resistivity of the cold plasma, the plasma current starts to decay (current quench) immediately. This causes the plasma to move continuously towards a new equilibrium state, leading to the overall vertical motion of the plasma~\cite{Kiramov:2018}. This is in contrast to a VDE caused by an inherently vertically unstable state of the plasma.

\begin{figure}[!h]
\subfigure[]{
\includegraphics[width=0.5\textwidth]{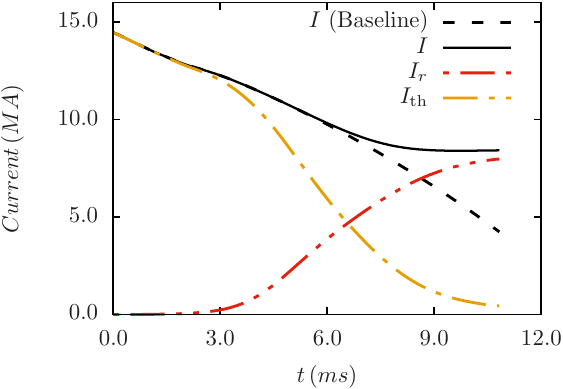}
\label{fig:current_decay}
}
\subfigure[]{
\includegraphics[width=0.5\textwidth]{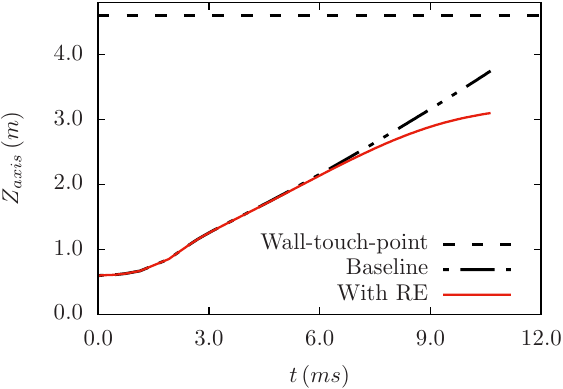}
\label{fig:zaxis_vs_time}
}
\caption{a) Evolution of plasma currents over time for the axisymmetric VDE simulation. Here, $I_\mathrm{th}$ and $I_{r}$ represent the thermal and runaway electron currents respectively. b) Plasma axis vertical position over time.}
\end{figure}

Figure~\ref{fig:current_decay} shows the plasma current decay and the simultaneous conversion of thermal current to RE current during the VDE. The decay is slowed down due to RE avalanching when a significant fraction of the current is carried by runaway electrons. Also the saturated total RE current is about $58\%$ of the predisruption current, which is in the range of the typically expected conversion ratio for ITER of about $\sim70\%$ \cite{Hender:2007}. The slowdown of the current decay leads to significant slowing down of the vertical plasma motion after about $\SI{7}{\milli\second}$, as shown in Fig.~\ref{fig:zaxis_vs_time}. This is due the fact that $Z_\mathrm{axis}$ is a pure function of the plasma current when the current quench time is much faster than the wall time, which is true in the present case.

\begin{figure}[!h]{}
\centering
\includegraphics[width=0.7\textwidth]{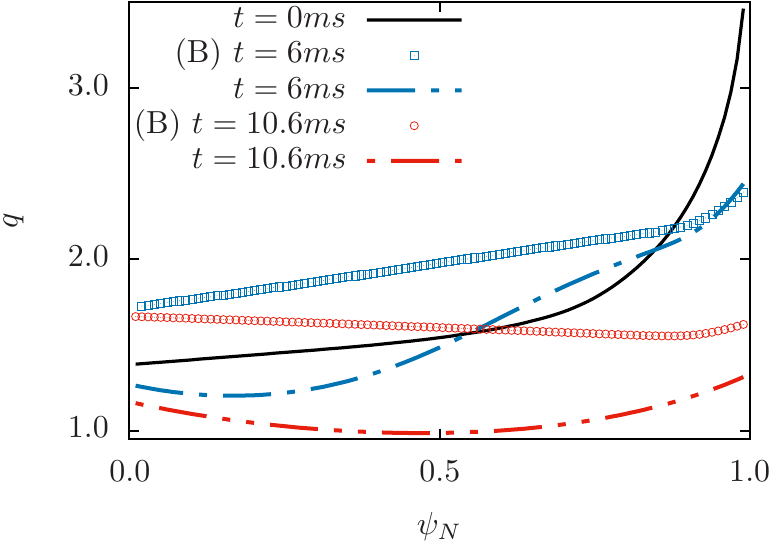}
\caption{Evolution of the q-profile over time for the axisymmetric VDE simulation with and without REs. The label (B) in the legend refers to the baseline case without REs.}
\label{fig:qprofiles}
\end{figure}

%\begin{figure}[!h]{}
%\centering
%\resizebox{0.6\textwidth}{!}{\input{j_profile_evolution_vs_rminor.tex}}
%\caption{ Evolution of the j-profile over time for the axisymmetric VDE simulation with and without REs. Solid lines refer to the case with REs.}
%\label{fig:jprofiles}
%\end{figure}

The corresponding evolution of q-profiles is shown in Fig.~\ref{fig:qprofiles}. We show that the conversion of thermal to RE current leads to significantly lower q-profiles. This is qualitatively similar to the observations from DINA simulations by Aleynikova \etal~\cite{Aleynikova:2016}. Due to the decay of the total current during the current quench phase, the q-profile near the center in general tends to rise. In the presence of REs, this effect is opposed by both the near-central peaking of the RE current profile as well as the reduced total current decay rate in the later phase of the RE conversion. \vinodh{The evolution of the edge safety factor $q_\mathrm{edge}$ is determined by two competing effects. The decay of the total plasma current tends to increase $q_\mathrm{edge}$, whereas the plasma scraping-off at the domain boundary tends to decrease $q_\mathrm{edge}$. The net effect however, is a decay of $q_\mathrm{edge}$ with time. This is due to the fact that the ratio $a^2/I$, which determines the approximate scaling of $q_\mathrm{edge}$ in an ideal circular plasma~\cite{Freidberg:2014}, decreases with time. Though the plasma that we consider here is far from ideal and circular, the qualitative picture remains the same}. The profile differences at time  $\SI{6}{\milli\second}$ \vinodh{are} purely due to RE current peaking, whereas the much lower q-profile with REs at  $\SI{10.6}{\milli\second}$ is both due to peaking and an overall higher total current. In our case, the peaking of the RE current profile is observed to be off-axis which suggests a longer timescale of parallel electric field diffusion\vinodh{~\cite{Eriksson:2004}} at the axis compared to the avalanche timescale. Values of $q$ lower than unity observed in this case can potentially destabilize the resistive internal kink mode.

\begin{figure}[!h]{}
\centering
\includegraphics[width=0.7\textwidth]{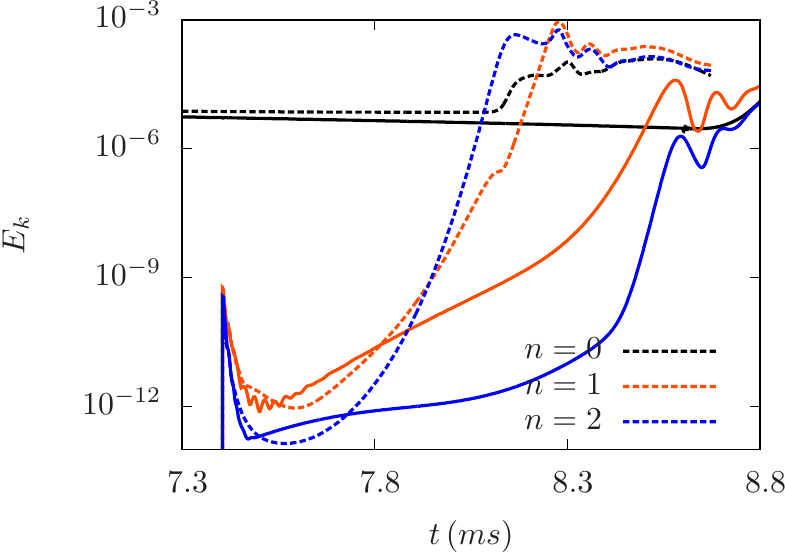}
\caption{Poloidal kinetic energy (in normalized units) of different toroidal modes integrated over the poloidal plane. Dotted lines refer to the case without runaway electrons and solid lines refer to the case with runaway electrons.}
\label{fig:pol_kin_energy}
\end{figure}

\begin{figure}[!h]
\subfigure[]{
\includegraphics[width=0.4\textwidth]{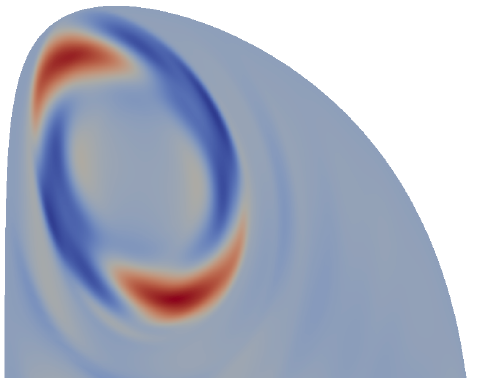}
}
\subfigure[]{
\includegraphics[width=0.4\textwidth]{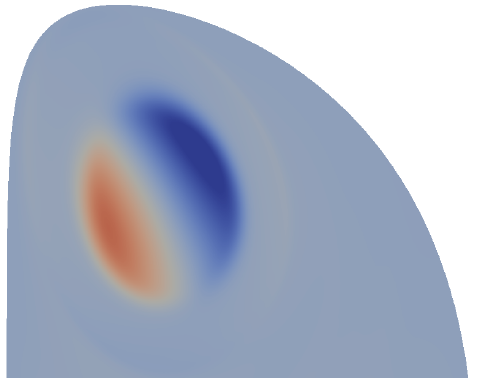}
}
\subfigure{
\includegraphics[width=0.12\textwidth]{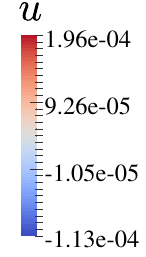}
}
\caption{$n=1$ and $n=2$ modes of the normalized electric potential $u$ during the non-linear phase of the mode growth at $t=\SI{8.6}{\milli \second}$. a) Without REs. b) With REs.}
\label{fig:mode_structure}
\end{figure}

We now turn to the non-axisymmetric simulations. In these cases, a very small perturbation in the axisymmetric state is applied at about $t=\SI{7.45}{\milli \second}$. In the case without RE generation, this leads to the exponential growth of the $n=1$ and $n=2$ resistive tearing modes, which eventually saturate at mode kinetic energies similar to the axisymmetric kinetic energy as shown in Fig.~\ref{fig:pol_kin_energy}. In the case with RE generation, $\left(m,n\right)=\left(2,1\right)$ is dominant in the initial phase of the mode growth. In addition, the $\left(m,n\right)=\left(2,1\right)$ mode grows slower due to the lower effective plasma resistivity with REs. Furthermore, the $\left(m,n\right)=\left(1,1\right)$ mode is observed to be eventually dominant in the case with RE generation compared to the $\left(m,n\right)=\left(2,1\right)$ mode which is dominant in the case without REs, as shown in Fig.~\ref{fig:mode_structure}. Such a qualitative behaviour is in agreement with the linear MHD analysis of Aleynikova \etal~\cite{Aleynikova:2016}. Similar behaviour is observed with the magnetic energies of the modes. 

\section{Summary and outlook}
\label{section5}

The runaway electron fluid model implemented in the non-linear MHD code JOREK has been presented. Being able to account for the back reaction from REs to the background plasma makes it a complementary tool to test-particle pusher codes \cite{Sommariva2:2018}. It is shown that Taylor-Galerkin numerical stabilization (TG2) enables RE advection velocities with timescales that are at least two orders of magnitude smaller than the timescale of MHD changes. This makes the code useful to study problems with RE-MHD interaction wherein the exact details/magnitude of parallel RE advection is insignificant. The ability to model RE parallel advection velocities close to the speed of light becomes important for understanding the deconfinement of REs in a stochastic field and the corresponding non-linear effects on MHD. To account for such scenarios, the fast parallel advection of REs in a stochastic field can be mimicked through a parallel diffusion term instead. The code was successfully benchmarked with the GO code\vinodh{~\cite{Papp:2013}} and its application to study the linear growth of the internal kink mode shows a recovery of the low resistivity scaling even at much higher resistivities as the fraction of RE current is increased. Furthermore, an axisymmetric cold VDE in an ITER plasma with simultaneous growth of REs was simulated using the model. Results show a significant slowing down of the vertical motion due to the formation of REs and the possibility of internal kink modes being destabilized due to $q$ values falling below unity. In addition, the presence of REs lead to a significantly different dynamics of \vinodh{the} 3D mode structure during the VDE.

We are currently exploring the application of the diffusion model for REs in JOREK to stochastic plasmas, which will be reported elsewhere. Furthermore, we intend to extend the numerical treatment of parallel diffusion in JOREK to even larger values of $D_{\parallel,r}/D_{\perp,r}$ through a scheme similar to the one presented in G{\"u}nter~\etal~\cite{Gunter:2007}, to avoid numerical pollution of perpendicular diffusion for large $D_{\parallel,r}$. Further plans include the implementation of the state-of-the-art treatment of RE generation in the presence of partially ionized impurities (incomplete screening) \cite{Hesslow:2018} and the development of more realistic hot-tail sources.

%% The Appendices part is started with the command \appendix;
%% appendix sections are then done as normal sections
%% \appendix

%% \section{}
%% \label{}

%% For citations use: 
%%       \citet{<label>} ==> Jones et al. [21]
%%       \citep{<label>} ==> [21]
%%

%cite{Helander:2007}

\section*{Acknowledgements}
VB and MH acknowledge fruitful discussions with P.~Helander, K.~Lackner, S.~G{\"u}nter, P.~Aleynikov and F.~Hindenlang. This work has been carried out within the framework of the EUROfusion Consortium and has received funding from the Euratom research and training programme 2014-2018 and 2019-2020 under the grant agreement No 633053. The views and opinions expressed herein do not necessarily reflect those of the European Commission. The related project number is AWP18-ERG-MPG/Bandaru. ITER is the Nuclear Facility INB no. 174. The views and opinions expressed herein do not necessarily reflect those of the ITER Organization. This publication is provided for scientific purposes only. Its contents should not be considered as commitments from the ITER Organization as a nuclear operator in the frame of the licensing process.

\section*{Appendix}

Normalization used for the variables that are encountered in the context of this paper are given the table below. The factors $\mu_0$, $n_0$, $\rho_0$ refer to the magnetic permeability of free space, central number density and the central mass density of the plasma respectively and $k_B$ is the Boltzmann constant.
\begin{center}
\begin{tabular}{|l|l|}
\hline
\textbf{Quantity} &  \textbf{Normalization} \\ \hline
Time, $t$ & $t^\mathrm{SI} = t \sqrt{\mu_0 \rho_0}$ \\ \hline
RE number density, $n_{r}$ & $n_{r}^\mathrm{SI} = n_{r} \sqrt{\rho_0/\mu_0} / \left(eR\right)$ \\ \hline
Speed of light, $c$ & $c^\mathrm{SI} = c/\sqrt{\mu_0 \rho_0}$ \\ \hline
Parallel electric field, $E_\parallel$ & $E_\parallel^\mathrm{SI} = E_\parallel/\sqrt{\mu_0 \rho_0}$ \\ \hline
Electric potential, $u$ & $u^\mathrm{SI} = u/ \sqrt{\mu_0 \rho_0}$ \\ \hline
Toroidal current density, $j$ & $j_\phi^\mathrm{SI} = -j/\left(R \mu_0 \right)$ \\ \hline
Resistivity, $\eta$ & $\eta^\mathrm{SI} = \eta \sqrt{\mu_0/\rho_0}$ \\ \hline
RE diffusivity, $D_{\perp,r}$ & $D_{\perp,r}^\mathrm{SI} = D_{\perp,r} /\sqrt{\mu_0\rho_0}$ \\ \hline
Density, $\rho$ & $\rho^\mathrm{SI} = \rho \rho_0$ \\ \hline
Temperature, $T$ & $T^\mathrm{SI} = T/\left(k_B \mu_0 n_0\right)$ \\ \hline
Parallel velocity, $v_{\parallel}$ & $v_{\parallel}^\mathrm{SI} = v_{\parallel} B/ \sqrt{\mu_0 \rho_0}$ \\ \hline

\hline
\end{tabular}
\end{center}
The RE primary and secondary source terms in normalized units are given below, along with the expressions for the constants that appear in them.
\begin{equation}
\begin{split}
 S_p &=  C_{1} R \rho^{(2-C_2)} T^{(C_2-3/2)}{|E_{\parallel}|}^{C_2}  
 \exp{\left[-C_3\frac{\rho}{|E_{\parallel}| T} - C_4 {\left(\frac{\rho}{|E_{\parallel}| T}\right)}^{\frac{1}{2}}\right]} 
 \notag \\& \times \exp{\left[-C_5{\frac{\rho^2}{{E_{\parallel}}^2 T}}  -  C_{6}{\frac{\rho^{3/2}}{{|E_{\parallel}|}^{3/2} T^{1/2}}} \right]}  
\end{split}
\end{equation} 
\begin{equation}
S_s =  C_{7} \rho n_{r} \left[C_8\frac{|E_{\parallel}|}{	\rho}-1\right] 
\left[1 - \frac{\rho}{C_8 |E_{\parallel}|} + \frac{C_{9} \rho^2}{\left({C_8}^2 {E_\parallel}^2+ C_{10} \rho^2\right)}\right]^{-\frac{1}{2}}
\end{equation}

\begin{align*}
C_d &= \frac{2\pi m_i^2}{c^4 e^3 \ln \Lambda \mu_o^{3/2} \rho_0^{1/2}},\\
 C_{00}& = \frac{m_i}{2c^2 m_e}, \quad C_{0} = {\left(1+Z_{e}\right)}^{1/2}, \\
C_1 &= \frac{\left(0.21 + 0.11Z_e\right) e^5 c^4 \ln \Lambda \mu_0^{5/2} \rho_0^{3/2} }{4\pi m_e^{1/2} m_i^{7/2}} {C_d}^{C_{2}},\\
C_{2}&=-\frac{3}{16}\left(1+Z_{e}\right),
\quad C_3= \frac{1}{4C_d},\quad C_4=\frac{C_{0}}{{C_d}^{1/2}}, \\
C_{5}&= \frac{C_{00}}{8 {C_d}^2} , \quad C_{6} = \frac{2}{3}\frac{C_{0}C_{00}}{{C_d}^{3/2}},
\quad C_{7}=\frac{c e^4\mu_0^{2} \rho_0}{4\pi m_e^2 m_i} \sqrt{\frac{\pi \varphi}{3\left(Z_e +5\right)}}, \\
 C_8 &= \frac{C_d}{C_{00}},\quad C_{9}=\frac{4\pi \left(Z_{e}+1\right)^2}{3\varphi\left(Z_{e}+5\right)},
\quad C_{10} =4/\varphi^2 -1.
\end{align*}

 \bibliographystyle{elsarticle-num} 
\bibliography{references_without_urls}

\end{document}